\documentclass{iaus} 
\usepackage{graphics} 
 
  \checkfont{eurm10} 
  \iffontfound 
    \IfFileExists{upmath.sty} 
      {\typeout{^^JFound AMS Euler Roman fonts on the system, 
                   using the 'upmath' package.^^J}%
       \usepackage{upmath}} 
      {\typeout{^^JFound AMS Euler Roman fonts on the system, but you 
                   dont seem to have the}%
       \typeout{'upmath' package installed. iaus.cls can take advantage 
                 of these fonts,^^Jif you use 'upmath' package.^^J}%
      } 
  \else 
  \fi 
 
  \checkfont{msam10} 
  \iffontfound 
    \IfFileExists{amssymb.sty} 
      {\typeout{^^JFound AMS Symbol fonts on the system, using the 
                'amssymb' package.^^J}%
       \usepackage{amssymb}%

      }{} 
  \fi 
 
  \IfFileExists{amsbsy.sty} 
    {\typeout{^^JFound the 'amsbsy' package on the system, using it.^^J}%
     \usepackage{amsbsy}} 
    {}

\title[The Interplay among Black Holes, Stars and ISM in Galactic  
       Nuclei]{Long-Term Profile Variability of Double-Peaked Emission Lines
in AGNs}

\author[Lewis {\it et al.\/}] 
{K. T. Lewis$^1$, M. Eracleous$^1$, J. P. Halpern$^2$
\& T. Storchi-Bergmann$^3$}
 
\affiliation{$^1$ Department of Astronomy and Astrophysics, 
The Pennsylvania State University, 525 Davey Laboratory, 
University Park, PA 16802, USA\\[\affilskip] 
$^2$ Department of Astronomy, Columbia University, 550 West 120th Street, 
New York, NY 10027, USA \\[\affilskip]
$^3$Instituto de Física, UFRGS, 91501-970 Porto Alegre, RS, Brazi } 
 
\pubyear{2004} 
\volume{222} 
\pagerange{1--2} 
\date{?? and in revised form ??} 
\jname{The Interplay among Black Holes, Stars and ISM \\in Galactic Nuclei} 
\editors{Th. Storchi Bergmann, L.C. Ho \& H.R. Schmitt, eds.} 
\begin{document} 
 
\maketitle 
 
\begin{abstract} 
An increasing number of AGNs exhibit broad, double-peaked Balmer
emission lines, which arise from the outer regions of the accretion
disk which fuels the AGN. The line profiles vary on timescales of
5--10 years. Our group has monitored a set of 20 double-peaked
emitters for the past 8 years (longer for some objects). Here we
describe a project to characterize the variability patterns of the
double-peaked H$\alpha$ line profiles and compare with those of two
simple models: a circular disk with a spiral arm and an elliptical
disk.
\end{abstract} 
 
\firstsection 
\section{Introduction}
Approximately 20\% of Broad-Line Radio Galaxies (BLRGs) exhibit broad,
double-peaked Balmer emission lines (Eracleous \& Halpern
2003). Double-peaked emission lines have also suddenly appeared in
LINERs, most notably NGC 1097 (Storchi-Bergmann et al. 1993). The
lines are thought to originate in the outer portions of the accretion
disk (R $\sim$ 100--10,000 r$_{g}$, where r$_{g} =
GM_{\bullet}/c^{2}$). The {\it profiles} of the double-peaked emission
lines are observed to vary on timescales of 5--10 yr and are a
manifestation of {\it physical} changes in the outer disk. Thus, the
profile variations of these lines can be used to test various models
for dynamical phenomena in the outer accretion disks of AGNs.

Here, we compare the data of PKS 0921--213 with two simple, but
physically motivated, models: a circular disk with a single precessing
spiral arm and a precessing elliptical disk. Spiral arms are present
in other astrophysical disks and provide a mechanism for removing
angular momentum from the disk. An elliptical disk could result from a
tidal perturbation of the disk by a massive object (e.g., a second
black hole) or from the tidal disruption of a star.

\textbf{Spiral Arm:} The spiral arm is modeled as a circular disk
with a perturbation in the emissivity pattern. The disk has the
following parameters: the inner and outer radii, $\xi_{1}$ and
$\xi_{2}$ (in units of $GM_{\bullet}/c^{2}$); the inclination angle
$i$; a broadening parameter $\sigma$; and the slope of the
axisymmetric emissivity law $q$ ($\epsilon \propto \xi^{-q}$). The
spiral arm perturbation is parameterized by: an amplitude $A$; pitch
angle $p$; width $\delta$; the inner radius of the arm $\xi_{sp}$; and
phase angle $\phi$.  \textbf{Elliptical Disk:} The elliptical disk
model is described by many similar parameters: $i$; $\sigma$; $q$;
$\phi$; $\xi_{1}$, and $\xi_{2}$, where $\xi$ is the {\it pericenter}
distance. The disk is circular up to $\xi_{e}$, then the eccentricity
increases linearly to a value of $e$ at $\xi_{2}$.

\section{Characterizing the Profile Variability} 
Both models involve many parameters and it is difficult to determine
which region of parameter space will be most successful in reproducing
the observed profile variations. Thus, as a first step, we have begun
characterizing the profile variability of our objects in a model
independent way. Each profile is reduced to a set of easily measured
quantities: the velocities of the red and blue peaks; the blue-to-red
peak flux ratio; the full widths at half and quarter maximum (FWHM and
FWQM); and the velocity shifts of the FWHM and FWQM centroids. Sets of
model profiles, with a variety of input parameters, can be
characterized in the same way for comparison with the data.

As an example, in Fig. 1 we show the profile parameter variations for
PKS 0921--213 (left), a spiral arm model (center) and an elliptical
disk model (right). The circular disk parameters ($\xi_{1}, \xi_{2}, q,
i,$ and $\sigma$) for both models are tuned to best-fit parameters of
the average profile of PKS 0931--213, while the additional model
parameters are simply chosen for illustration. The most striking
difference between the two models is that the elliptical disk profiles
{\it always} vary smoothly with phase, while those of the spiral arm
model vary much more sharply. However, both models show potential for
reproducing the observed profile variation. By creating ``libraries''
of model profiles, we will be able to quickly select a small region
of parameter space to test.

\begin{figure} 
 \includegraphics{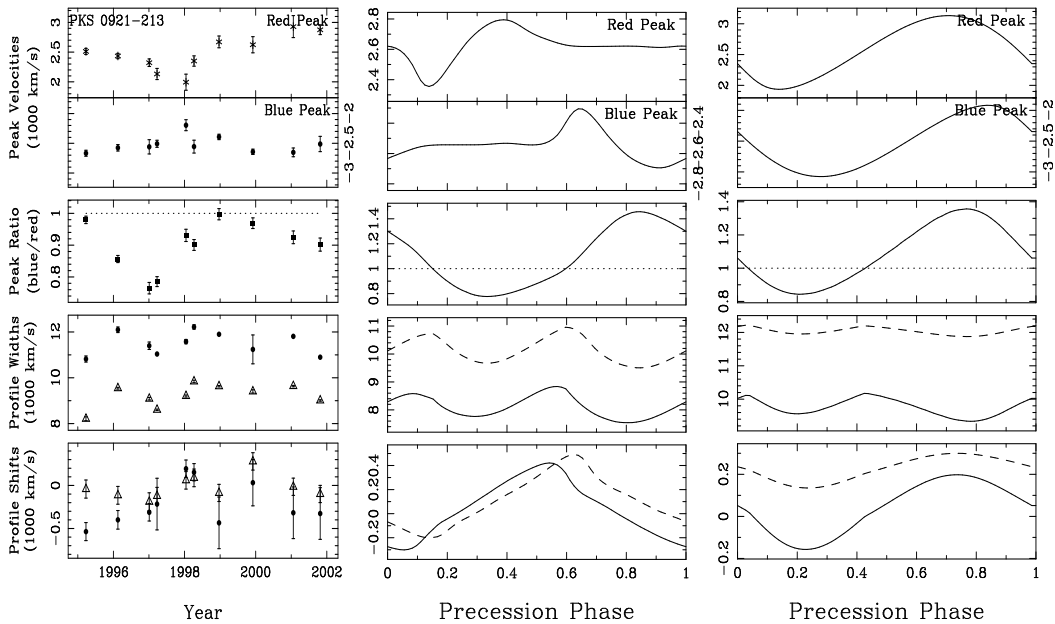} \caption{Variability of data and model
 profile properties with time. ({\it Left}) \textbf{PKS 0921--213:}
 The FWHM and FWQM, and corresponding shifts, are denoted with
 triangles and filled circles, respectively. ({\it Center})
 \textbf{Spiral Arm Model:} Variation of profile properties with phase
 for a disk with a one-armed spiral. The model parameters are:
 $q$=1.5; $i$=50$^{\circ}$; $\xi$=1200--8000r$_{g}$; $\sigma$=600
 km/s; $A$=5; p=15$^{\circ}$; $\delta$=35$^{\circ}$; and
 $\xi_{sp}$=1200r$_{g}$. The FWHM and FWQM are denoted with solid and
 dashed lines, respectively. ({\it Right}) \textbf{Elliptical Model:}
 Same as the {\it Center} panel, but for an elliptical disk with
 $q$=1.5, $i$=50$^{\circ}$, $\xi$=1200--8000 r$_{g}$, $\sigma$=600
 km/s, $e$=0.15, and $\xi_{e}$=1200r$_{g}$.}
\end{figure}

\begin{acknowledgments} 
KTL is supported by a NASA GRSP grant NGT5-50387.
\end{acknowledgments}


\begin{thebibliography}{} 
 
  \bibitem[Eracleous \& Halpern (2003)]{EH94} 
     {Eracleous, M.~ \& Halpern, J.~P.} 2003 
     \textit{ApJ} \textbf{599}, 886. 

 \bibitem[Storchi-Bergmann et al.(1993]{SB93} 
     {Storchi-Bergmann, T., Baldwin, J.~A., \& Wilson, A.~S.} 1993 
     \textit{ApJL} \textbf{410}, L11. 
 
\end{thebibliography}
\end{document}